\documentclass[9pt,review]{elsarticle}

\usepackage{hyperref}

\usepackage{amssymb,amsmath,bm,mathrsfs,amsfonts,graphicx,multirow,stmaryrd}

\usepackage{times}
\usepackage{natbib,dsfont}
\usepackage{amsthm}
\usepackage{color}
\usepackage{enumitem}
\usepackage{amssymb,amsmath,bm,mathrsfs,amsfonts,graphicx,multirow,stmaryrd,color}
\usepackage{amsmath,amssymb,amsthm,enumerate,epsfig,graphicx}
\usepackage{amsmath,amsthm,amssymb,amscd,txfonts}
\usepackage{times}
\usepackage{natbib,dsfont}
\usepackage{amsthm}
\usepackage{epstopdf}
\usepackage{graphics}
\usepackage{subfig}
\usepackage{float}
\usepackage{mathrsfs}
\usepackage{booktabs}
\usepackage{footnote}
\usepackage{graphicx}
\usepackage{epstopdf}
\usepackage{caption}
\usepackage{enumerate}
\usepackage{soul}
\usepackage{color,xcolor}

\numberwithin{equation}{section}
\theoremstyle{plain}
\theoremstyle{remark}

\newtheorem{theorem}{Theorem}

\newcommand{\lam}{{\lamega}}

\newcommand{\bmu}{{\bm \mu}}

\newcommand{\bSig}{{\bm \Sigma}}

\newcommand{\bqa}{\begin{eqnarray}}
\newcommand{\eqa}{\end{eqnarray}}
\newcommand{\bqn}{\begin{eqnarray*}}
	\newcommand{\eqn}{\end{eqnarray*}}
\newcommand{\be}{\begin{equation}}
\newcommand{\ee}{\end{equation}}
\allowdisplaybreaks[4]
\numberwithin{equation}{section}

\renewcommand{\(}{\left(}
\renewcommand{\)}{\right)}

\def\lam{\lamega}
\def\to{\rightarrow}

\def\vp{{\varphi}}

\def\phi{\varphi}
\def\lam{{\lambda}}

 \def\gam{\gamma}

\newtheorem{thm}{Theorem}[section]

\theoremstyle{definition}

\numberwithin{equation}{section}
\numberwithin{thm}{section}

\theoremstyle{remark}
\newtheorem{assumption}[thm]{Assumption}

\def\fX {{\mathbf X}}
\def\fY {{\mathbf Y}}
\def\fx {{\mathbf x}}

\def\tfC {{\widetilde{\mathbf C}}}

\def\fv {{\mathbf v}}
\def\fu {{\mathbf u}}
\def\fw {{\mathbf w}}
\def\fS {{\mathbf S}}

\def\fr {{\mathbf r}}

\def\fc {{\mathbf c}}

\def\fI {{\mathbf I}}

\def\funit {{\mathbf 1}}

\def\fy {{\mathbf y}}

\newcommand{\beq}{\begin{eqnarray}}
\newcommand{\eeq}{\end{eqnarray}}
\newcommand{\beqq}{\begin{eqnarray*}}
\newcommand{\eeqq}{\end{eqnarray*}}

\journal{Journal of \LaTeX\ Templates}

\begin{document}
	
\begin{frontmatter}	
	
\title{Spectrally-Corrected and Regularized Global Minimum Variance Portfolio for Spiked Model}	

\author[mymainaddress]{Hua Li\corref{mycorrespondingauthor}}\ead{lihua@ccu.edu.cn}

\author[mymainaddress]{Jiafu Huang}


\cortext[mycorrespondingauthor]{Corresponding author}

\address[mymainaddress]{School of Science, Chang Chun University, China}

\begin{abstract}
Considering the shortcomings of the traditional sample covariance matrix estimation, this paper proposes an improved global minimum variance portfolio model and named spectral corrected and regularized global minimum variance portfolio (SCRGMVP), which is better than the traditional risk model. The key of this method is that under the assumption that the population covariance matrix follows the spiked model and the method combines the design idea of the sample spectrally-corrected covariance matrix and regularized. The simulation of real and synthetic data shows that our method is not only better than the performance of traditional sample covariance matrix estimation (SCME), Ledio-Wolf's shrinkage estimation (SHRE), weighteid shrinkage estimation (WSHRE) and simple spectral correction estimation (SCE), but also has lower computational complexity.
\end{abstract}	

\end{frontmatter}


\section{Introduction}
\indent Due to the development of the financial market, investment funds have increased significantly. Therefore, investors must know the proportion of the portfolio to maximize benefits and reduce the risks associated with investment decisions. The mean-variance portfolio proposed by Markowitz \cite{1952Portfolio, 1959Portfolio} focuses primarily on determining the optimal portfolio weight, i.e. the proportion of wealth invested in financially risky assets. It is one of the key components of capital market theory in dealing with financial investment decisions. By this theory, investors can construct an optimal portfolio by minimizing the portfolio variance at a given expected return or maximizing the portfolio return at a given portfolio risk. One of Markowitz's ideas was the minimization of the portfolio variance subject to the budget constraint, which is called the global minimum variance portfolio(GMVP); see arguments in  \cite{1993The,2019A}. The GMVP is the most commonly used and efficient investment decision by researchers in the financial field. This portfolio has the smallest variance of all optimal portfolios as the solution to the Markowitz mean-variance optimization problem.

First, we would like to recall the GMVP. We have a financial market consisting of M risky assets, and let $\fy_t$ be the M-dimensional time series of returns at time $t=1,\cdots,n$. Thus, we have the following random vector:
\beqq
\fy_t&=&\(y_{1t},y_{2t},\cdots,y_{Mt}\)^T. 
\eeqq
where       $\fy_t\sim{N(\bmu_t,\bSig_t)}$. 
Then,  we typically model the following vector stochastic processes based on expected and unexpected reward contributions:
\beq
\fy_t&=&\bmu_t+\bSig_t^\frac{1}{2}\fx_t.
\eeq
where $\bmu_t$ and $\bSig_t=(\sigma_{ij})$ are the expected value and covariance matrix of the asset returns over an investment period, and $\fx_t $ is a random vector with independent and identically distributed entries having mean zero and variance one, we care about the weight corresponding to the minimum variance portfolio. Let $ \fw_t=\(w_{1t},w_{2t},...,w_{Mt}\)^T $ denotes the vector of portfolio weights, the expected value and covariance matrix of the portfolio are $ E(r_t)=\fw_t^T\bmu_t ,Var(r_t)=\fw_t^T\bSig_t\fw_t$ respectively, where the $ \fr_t=\(r_{1t},r_{2t},...,r_{Mt}\)^T $ is random returns. During a single investment period in GMVP, under the constraint of $\fw^T\funit_M=1$, the risk of the minimum variance portfolio is expressed as $\sigma_M^2(\fw)=\fw^T\bSig\fw$. Accordingly, the global minimum variance portfolio selection problem can be formulated as the quadratic optimization problem with liner constraints, which is given by :
\beqq
&\mbox{min: }&\sigma_M^2(\fw)=\fw^T\bSig\fw,\\
&\mbox{s.t. }& \fw^T\funit_M=1.
\eeqq
where $\funit_M=(1,...,1)^T$ stands for M dimensional vector of ones, and $\fw$ denotes the vector of portfolio weights. The solution to the former optimization problem is given by:
\beq\label{weight}
\fw_{GMVP}&=&\frac{\bSig^{-1}\funit_M}{\funit_M^T\bSig^{-1}\funit_M}.
\eeq
The global minimum variance portfolio \ref{weight} has the smallest variance of all portfolios.
Then, the global minimum variance can be given by:
\beq
\sigma^2(\fw_{GMVP})&=&\frac{1}{\funit_M^T\bSig^{-1}\funit_M}.
\eeq

The weights of GMVP have several desirable properties and have good applicability in practice. However, the population covariance matrix $\bSig$ is unknown and so it must be estimated using historical data of the asset returns. The quality of the estimator $\bSig$  has a strong influence on GMVP. Therefore, the estimation of GMVP is closely related to the estimation of the population covariance matrix for asset returns.

Traditional estimators are a commonly used method for the estimation of the GMVP (\ref{weight}). The traditional method is to replace the population covariance matrix $\bSig$ with a sample covariance $\fS$.
\beq
\fS=\frac{1}{n-1}\sum_{i=1}^{n}(\fy_i-\overline{\fy} )(\fy_i-\overline{\fy} )^T, 
\eeq
where $\overline{\fy}=\frac{1}{n}\sum_{i=1}^{n}y_i$. Then, the classic estimator of the $\fw_{GMVP}$ can be expressed as:

\beqq
\hat{\fw}_S&=&\frac{\fS^{-1}\funit_M}{\funit_M^T\fS^{-1}\funit_M}.
\eeqq

In the classical asymptotic analysis, it is generally assumed that the number M of the investment asset is fixed and the sample size n is increased. In this case, the sample estimator is a consistent estimator and is often called standard asymptotics \cite{2011Optimizing,Yarema2006Distributional,2007A}. However, when both the asset dimension M and the sample size n tend to infinity, or more specifically, M and n have a considerable size ($M/n=J>0$) classical estimator effect is not ideal, and even the consistency of the estimator is not guaranteed. Thus, it is inappropriate to use the sample covariance matrix to build a high-dimensional portfolio. To improve the performance of investment portfolios, two different methods are very popular. The first method is to directly constrain \cite{2019A,0Vast,Patrick2013On,2015Sparse} or shrink \cite{2007Multivariate,2014Estimation} the portfolio weights. The second method is to use the improved covariance matrix estimator to constrain or shrink the portfolio weights, such as  \cite{2018improved,2016Efficient}. Both dimension M and sample size n tend to infinity, which is known as high-dimensional asymptotics in \cite{bzd2006Spectral, bao2022spectral}. In this case, new asymptotic techniques must be applied for derivations. Therefore, many articles have studied the characteristics of high-dimensional global minimum variance investment portfolios, such as \cite{2020High,2015Do,0Vast}.

There are many articles provide different methods to improve the estimation of high-dimensional covariance matrix, which can be divided into two categories. The first category is based on additional knowledge in the estimation process, such as sparseness, graph models, and factor models \cite{Rajaratnam2008Flexible,2011ESTIMATION,2008High,2012Minimax}. The second category is the spectrum of corrected sample covariance matrices, such as Ledoit and Wolf's (linear and nonlinear) shrinkage estimates \cite{2004A,2011Nonlinear,2010Nonlinear} and other shrinkage estimators for covariance matrices \cite{2001Shrinkage}. In this paper, we proposed method belongs to the second category, and it integrates the design ideas of the sample spectrally-corrected covariance matrix \cite{2013The} and the regularized method \cite{2008Regularized} to improve the GMVP estimation in high-dimensional settings. The details are given in the following sections.

In this paper, an extremely important and widely used model is involved, which is known as the spiked model in \cite{ONATSKI2014DETECTION}. The model assumes that the population covariance matrix has only a fixed number of eigenvalues that are not equal to one, and all other eigenvalues are equal to one, and names those eigenvalues that are not one as "spikes". This model has been used in many real applications, we learned about mathematical finance \cite{2008Determining}, EEG signals \cite{2009Functional,20111}, and data analysis \cite{2004Limiting}. The most important result of the spiked model is the study of eigenvalues of the sample covariance matrix, such as \cite{2021Estimation,2006Eigenvalues,1998No,2004CLT}.  According to the related theory of the spiked covariance model, the population covariance matrix can be estimated by a class of sample covariance matrices that follow the structure of the spiked model. That is, the sample covariance matrix is written as a finite rank perturbation of a scaled identity matrix. The principal eigenvectors of the sample covariance matrix give the direction of the low-rank perturbation, correct the corresponding eigenvalues to one, and regularized by adding appropriate parameters.

According to the random matrix theory, derive the approximation of the portfolio variance, the regularization parameters are some of the parameters we choose to minimize the approximation of the portfolio variance. In this way, we not only preserve the structure of the spiked model, but also reduces the number of design parameters.

The rest of this paper is organized as follows: Section \ref{mainresult} shows the spectrally-corrected and regularized estimates of the covariance matrix.
In section \ref{estimate value}, provide a consistent estimate of our proposed method (SCRGMVP) and describe the optimal parameters. The performance of our method is studied in section \ref{simulation study} using synthetic and real simulations before concluding in section \ref{conclu}.

\section{SCRGMVP under Spiked Model}\label{mainresult}

In this paper, we propose a particular form of the covariance matrix, wherein $\bSig$ takes the following form:
\begin{eqnarray}\label{pcm}
	\bSig&=&\sigma^2\left(\fI_M+\sum_{j\in \mathbb{I}_1}\lambda_{j}\fv_j\fv_j^T+\sum_{j\in \mathbb{I}_2}\lambda_{j}\fv_{j}\fv_{j}^T\right).
\end{eqnarray}
where $\sigma^2>0$, $\mathbb{I}_1=\left\lbrace 1,\cdots,r_1 \right\rbrace$, $\mathbb{I}_2=\left\lbrace -r_2,\cdots,-1 \right\rbrace$, $r=r_1+r_2$, $\lambda_1\geq\cdots\geq\lambda_{r_1}>0>\lambda_{-r_2}
\geq\cdots\geq\lambda_{-1}>-1$ and $\fv_1,\cdots,\fv_{r_1},\fv_{-r_2},\cdots,\fv_{-1}$ are orthonormal. This is the spiked model, which is used in many real applications, such as mathematical finance, EEG signals, and mathematical analysis. In practice, we can estimate parameters through algorithms in many relevant literature, such as \cite{2012ESTIMATION,2021Generalized,2021Estimation,2020Estimating}. Therefore, we can assume that the parameters $\sigma^2$, $r_1$, $r_2$, and $\lambda_i$($i\in \mathbb{I}$) are completely known.

Based on \ref{pcm}, we consider a class of sample covariance matrix estimators that follow the same spike model, that is, the sample covariance matrix is written in the finite rank perturbation form of the identity matrix. First, we start from the eigendecomposition of the sample covariance matrix:
\beq
\fS = \sum_{i=1}^Ms_j\fu_j\fu_j^T
\eeq
with $s_j$ being the $j$-th largest eigenvalue and $\fu_j$ its corresponding eigenvector, and correcting $s_j$ to the corresponding one of $\Sigma$ as follows:
\begin{eqnarray}\label{eq:tfS}
	\hat{\bSig}&=&\sigma^2\left(\fI_M+\sum_{j\in \mathbb{I}_1}\lam_{j}\fu_j\fu_j^T+\sum_{j\in \mathbb{I}_2}\lam_{j}\fu_{j}\fu_{j}^T\right).
\end{eqnarray}
Comparing \ref{pcm} and \ref{eq:tfS}, we know that $\hat{\bSig}$ is the same as $\Sigma$ for the spiked model except for the spiked eigenvectors. In this way, the structure of $\Sigma$ is preserved as much as possible. However, it is well known that $\hat{\bSig}$ is a biased estimate of the population covariance matrix $\Sigma$. Accordingly, it is necessary to add  regularization parameters to the sample spiked covariance matrix. So we can get the estimation of the global minimum variance portfolio weight and it is expressed by: 
\beq
\sqrt{M}\hat{\fw}_{SCRGMVP}&=&\frac{\tfC \frac{\funit_M}{\sqrt{M}}}{\frac{\funit_M^T}{\sqrt{M}} \tfC \frac{\funit_M}{\sqrt{M}}},
\eeq
where
\begin{eqnarray}\label{eq:tfH}
	\tfC&=&\left(\fI_M+\gamma_1\sum_{j\in \mathbb{I}_1}\lam_{j}\fu_j\fu_j^T+\gamma_2\sum_{j\in \mathbb{I}_2}\lam_{j}\fu_{j}\fu_{j}^T\right)^{-1}.
\end{eqnarray}
where $\gamma_1$ and $\gamma_2$ are the regularization parameters that need to be optimized. 
Therefore, the spectrally-corrected and regularized global minimum variance portfolio risk  ($\sigma_M^2(\hat{\fw}_{SCRGMVP})$) is expressed as:

\begin{align}
M\sigma_M^2(\hat{\fw}_{SCRGMVP})&=\frac{\frac{\funit_M^T}{\sqrt{M}}\tfC\bSig\tfC\frac{\funit_M}{\sqrt{M}}}{\(\frac{\funit_M^T}{\sqrt{M}}\tfC\frac{\funit_M}{\sqrt{M}}\)^2}\notag\\
&=\frac{\fY(\tfC,\bSig)}{\(\fX(\tfC)\)^2},
\end{align}
in which
\beq\label{eq:y}
\fY\(\tfC,\bSig\)&=&\frac{\funit_M^T}{\sqrt{M}}\tfC\bSig\tfC\frac{\funit_M}{\sqrt{M}},
\eeq

\beq
\fX\(\tfC\)&=&\frac{\funit_M^T}{\sqrt{M}}\tfC \frac{\funit_M}{\sqrt{M}}.
\eeq

\section{Consistent Estimator of the Portfolio Risk and Parameter Optimization}\label{estimate value}
In this section, the main task is to select an optimal set of parameters $\gamma=\{\gamma_1,\gamma_2\}$ to minimize the risk of SCRGMVP:
\beq\label{eq:min0}
\gamma^*=\mathop{\arg \min}\limits_{\gamma>0}\(M\sigma^2_{SCRGMVP}(\gamma)\)
\eeq
where $\gamma^*=\{\gamma_1^*,\gamma_2^*\}$.
It is usually not possible to obtain an exact optimal solution $\gamma$ for \ref{eq:min0}. To solve this problem, we need to approximate the risk of SCRGMVP in the asymptotic growth state defined in the following assumption based on the results of random matrix theory. The parameters corresponding to the minimum approximation are the theoretical optimal parameters $\gamma_1^*, \gamma_2^*$.

\begin{assumption}\label{as:1}
$M,n\rightarrow \infty$ and the following limits exist: $\frac{M}{n}\rightarrow J>0$, and $\frac{M}{n}\rightarrow J<1$.
\end{assumption}
Assumption \ref{as:1} is usually a key assumption under high-dimensional random theory.

\begin{assumption}\label{as:2}
$r_1$ and $r_2$ are fixed and {{$\lam_1>\cdots>\lam_{r_1}>\sqrt{J}>0>-\sqrt{J}>\lam_{-r_2}>\cdots>\lam_{-1}>-1$}}, independently of $M$ and $r=r_1+r_2$.
\end{assumption}

Assumption \ref{as:2} is the basis for our analysis results, ensuring a one-to-one mapping between sample eigenvalues $s_j$ and unknown population eigenvalues $\lam_j$. In fact, when $\lam_j>\sqrt{J}
(j \in \mathbb{I}_1)$ or $-1<\lam_{j}<-\sqrt{J} (j \in \mathbb{I}_2)$, $\lam_j$ can be consistently estimated using its corresponding $s_j$ for $j \in \mathbb{I}$. Otherwise, the relation between $s_j$ and $\lam_j$ no longer holds.

\begin{assumption}\label{as:3}
$\|\bmu\|$ has a bounded Euclidean norm, that is $\|\bmu\|=O(1)$.
\end{assumption}

\begin{assumption}\label{as:4}
The spectral norm of $\Sigma$ is bounded, that is $\|\bSig\|=O(1)$.
\end{assumption}

\begin{theorem}\label{thm:1}
Under Assumptions \ref{as:1} to \ref{as:4}, we have
\beqq
&&\fX\left(\tfC\right)-\overline{X}(\gamma_1,\gamma_2)\xrightarrow{a.s.} 0,
\\
&&\fY\left(\tfC,\bSig\right)-\overline{Y}(\gamma_1,\gamma_2) \xrightarrow{a.s.} 0,
\eeqq

where
\beq
\overline{X}\left(\gamma_1,\gamma_2\right)
&=&1-\sum_{i=1,2}\sum_{j\in \mathbb{I}}a_jb_j\gamma_{i,j}\delta_{i,j}\label{eq:oG}\\
\overline{Y}(\gamma_1,\gamma_2)
&=&1+\sum_{i=1,2}\sum_{j\in \mathbb{I}}\lam_jb_j\delta_{i,j}
-2\sum_{i=1,2}\sum_{j\in \mathbb{I}}\(\lam_j+1\)a_jb_j\gamma_{i,j}\delta_{i,j}\nonumber\\
&&+\sum_{i=1,2}\sum_{j\in \mathbb{I}}a_jb_j\(\lam_ja_j+1\)\gamma_{i,j}^2\delta_{i,j}\label{eq:oD}
\eeq
in which
\beqq
\gamma_{i,j}=\frac{\gamma_i\lam_j}{1+\gamma_i\lam_j},\;\;a_j = \frac{\lam_j^2-J}{\lam_j(\lam_j+J)},\;\;b_j = \(\frac{\funit_M^T}{\sqrt{M}}\fv_j\)^2,\notag
\eeqq
for $i=1,2$ and $j\in \mathbb{I}$ and

\end{theorem}

\begin{theorem}
According to theorem \ref{thm:1}, the deterministic equivalent of the global minimum variance portfolio can be expressed as :
\begin{align*}
	M\sigma_M^2\(\hat{\fw}_{SCRGMVP}\)-M\overline{\sigma}_M^2\(\hat{\fw}_{SCRGMVP}\)\xrightarrow{a.s.} 0
\end{align*}
where
\begin{align}
	M\overline{\sigma}_M^2(\hat{\fw}_{SCRGMVP})&=
	\frac{\overline{Y}\left(\gamma_1,\gamma_2\right)}{\( \overline{X}\left(\gamma_1,\gamma_2\right)\)^2}
\end{align}

\end{theorem}
\begin{theorem}\label{thm:3}
	Under the setting of Assumption 1, the optimal parameters ${\gamma_1^*,\gamma_2^*}$ that minimize $\overline{\sigma^2}_{SCRGMVP}(\gamma_1,\gamma_2)$ are given by:
	\beqq
	\gamma_1^* = \frac{\vp_1^*}{(1-\vp_1^*)\lam_1}, \;\;\mbox{and}\;\;\gamma_2^* = \frac{-\vp_2^*}{(1-\vp_2^*)\lam_{-1}}.
	\eeqq
	where $(\vp_1^*,\vp_2^*)$ is the minimizer of the function $g(\vp_1,\vp_2)$, $\vp_{i}\in{(0,1)},i=1,2$ given by
	\beqq
	g(\vp_1,\vp_2)=\frac{\overline{Y}\left(\vp_1,\vp_2\right)}{\( \overline{X}\left(\vp_1,\vp_2\right)\)^2}
	\eeqq
	where
	\beqq
	\overline{X}\left(\vp_1,\vp_2\right)
	&=&1-\sum_{i=1,2}\sum_{j\in \mathbb{I}}a_jb_j\gamma_{i,j}\delta_{i,j}\\
	\overline{Y}(\vp_1,\vp_2)
	&=&1+ \sum_{j\in \mathbb{I}}\lam_jb_j
	-2\sum_{i=1,2}\sum_{j\in \mathbb{I}}\(\lam_j+1\)a_jb_j\gamma_{i,j}\delta_{i,j}\\
	&&+\sum_{i=1,2}\sum_{j\in \mathbb{I}}a_jb_j\(\lam_ja_j+1\)\gamma_{i,j}^2\delta_{i,j}\\
	\gamma_{1,j}&=&\frac{\vp_1\lam_j}{(1-\vp_1)\lam_1+\vp_1\lam_j},\;\;\;j \in \mathbb{I}_1,\\
	\gamma_{2,j}&=&\frac{-\vp_2\lam_{j}}{(1-\vp_2)\lam_{-1}-\vp_2\lam_{j}},\;\;\;j \in \mathbb{I}_2.
	\eeqq
	
\end{theorem}
\begin{theorem}\label{th:lrmt}
Under Assumptions \ref{as:1} to \ref{as:4} and fix an $i\in \mathbb{I} = \{-r_2,\dots, -1,1,2,\dots,r_1\}$, for any deterministic unit vector $\fc\in S_{\mathbb{R}}^{M-1}$, we have
\begin{equation}\label{eq:angle}
\left\langle \fc, \fu_i\fu_i^T\fc \right\rangle \to \dfrac{\lam_i^2-J}{\lam_i(\lam_i+J)}\left\langle \fc, \fv_i\fv_i^T\fc \right\rangle ,
\end{equation}
almost surely. This theorem has been proven in \cite{2022Spectrally}, so it is omitted here.
\end{theorem}

\begin{theorem}\label{thm:6}
	Under Assumption 1, we have
	$$\left|\lam_j-\hat{\lam}_j\right|\stackrel{a.s,} {\longrightarrow}{0},\quad
	\left|b_j-\hat{b}_j\right|\stackrel{a.s,} {\longrightarrow}{0},$$
	where
	\beq \label{hat_lammda}
	\hat{\lam}_j=\frac{s_j/{\sigma^2}+1-J+\sqrt{\(s_j/{\sigma^2}+1-J\)^2-4s_j/{\sigma^2}}}{2}-1,
\eeq
\beq \label{hat_bj}
	\hat{b}_j=\frac{1+J/\hat{\lam}_j}{1-J/\hat{\lam}_j}
\frac{\frac{\funit_M^T}{\sqrt{M}}\fu_j\fu_j^T\frac{\funit_M}{\sqrt{M}}}
{1-J\sigma^2},
\eeq
where, $J=\frac{M}{n}$, $s_j$ is the jth maximum eigenvalue of the sample covariance matrix, and $j \in \mathbb{I}$.
\end{theorem}

\section{Simulation}\label{simulation study}
This section is mainly divided into three parts. The first part mainly compares whether the changing trends of $\sigma_M^2\(\hat{\fw}_{SCRGMVP}\)$ and $\overline{\sigma}_M^2\(\hat{\fw}_{SCRGMVP}\)$ are consistent; The second and third parts respectively compare our proposed method with other classic methods using synthetic data and multiple real stock data.

\subsection{Consistent Estimation Comparison}
In this paper, due to the optimization of parameter $\sigma_p^2\(\hat{\fw}_{SCRGMVP}\)$, we need to use the relevant results of random matrix theory to derive the consistent estimation $\overline{\sigma}_p^2\(\hat{\fw}_{SCRGMVP}\)$ of risk $\sigma_p^2\(\hat{\fw}_{SCRGMVP}\)$ in the large dimension environment of our proposed method. To verify the correctness of our derived consistent estimation, we need to compare whether the changing trends of $\sigma_p^2\(\hat{\fw}_{SCRGMVP}\)$ and $\overline{\sigma}_p^2\(\hat{\fw}_{SCRGMVP}\)$ are consistent.

Next, we describe the commonly used simulation settings for experiments using synthetic data.Specifically, we set $\sigma^2=1$ and choose $r_1=3$, $r_2=1$ . Provide the eigenvalues $\lam_{1}=20$, $\lam_{2}=10$, $\lam_3=5$, $\lam_M=0.01$, and set the corresponding eigenvectors as follows: $v_1=(1,0,\cdots,0)^{'}$, $v_2=(0,1,0,\cdots,0)^{'}$, $v_3=(0,0,1,\cdots,0)^{'}$,  $v_{M}=(0,\cdots,0,1)^{'}$. Construct the covariance matrix of the spiked model structure according to the above settings. Using the training set, design the improved GMVP method using grid search. The performance of this risk and its estimation is illustrated in Figure 1, we can see how our asymptotic deterministic equivalence accurately approximates the true variance, gradually converging to the true variance as the sample size increases. Therefore, in high-dimensional situations, we can use grid search to find the optimal parameter $\gamma^*$.

\begin{figure}[H]
	\setlength{\abovecaptionskip}{0pt}
	\setlength{\belowcaptionskip}{0pt}
	\centering
		\includegraphics[height=8cm,width=12cm]{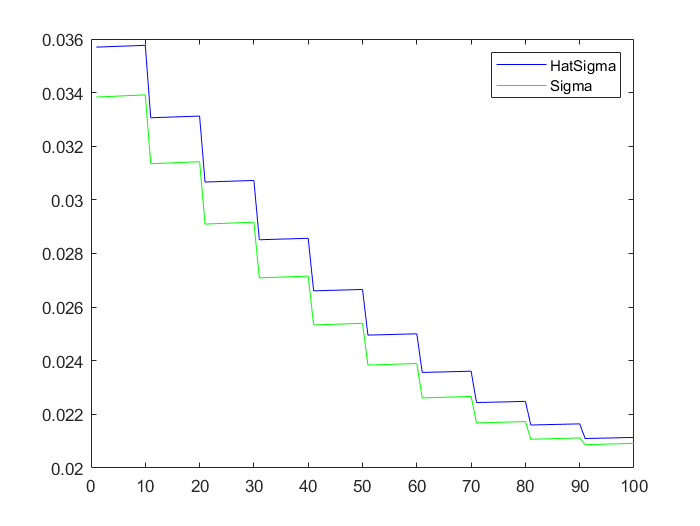}
	\caption{Portfolio variance vs. sample size $n$ for $M=50$. Comparison of SCRGMVP(Sigma)and its consistent estimation(HatSigma) with synthetic data.}
	\label{fig2}
\end{figure}

\subsection{Synthetic Data Simulation}

In this section, we evaluated the approximate accuracy of the following five different estimates: 1) the classical estimate, namely the sample covariance matrix estimator (SCME); 2) we propose the spectrally-corrected and regularized covariance matrix estimator (SCRE); 3) replace the unknown population covariance matrix with the shrinkage covariance matrix estimator(SHRE) proposed by Ledio-Wolf in (2004); 4) The two other methods are ordinary spectral correction estimator(SCE) and the weighted shrinkage estimator(WSHRE). We use Monte Carlo simulation methods to estimate portfolio risk and the specific steps are as follows:
\begin{itemize}
	\item Step 1: Set $\sigma^2=1$ and choose $r_1=3$, $r_2=1$ , orthogonal vectors as follows: $v_1=(1,0,\cdots,0)^{'}$, $v_2=(0,1,0,\cdots,0)^{'}$, $v_3=(0,0,1,0,\cdots,0)^{'}$,  $v_{M}=(0,\cdots,0,1)^{'}$ and their corresponding weights $\lam_{1}=20$, $\lam_{2}=10$, $\lam_3=5$, $\lam_M=0.01$. Randomly generate p-dimensional mean vectors in intervals $\(-1,1\)$.
	\item Step 2: Generate $n$ training samples.
	\item Step 3: Derive the spectral correction and the optimal parameters $\gamma^{*}_1$, $\gamma^{*}_2$ of the sample covariance of the SCRGMVP method using grid search over $\{\vp_1,\vp_2\}\in{\{[0,1)\times{[0,1)}\}}$.
	\item Step 4: Calculate the portfolio risks corresponding to these five population covariance estimates.
	\item Step 5: Repeat steps 2-4 10000 times and calculate the average portfolio risk corresponding to these five methods.
\end{itemize}

\begin{figure}[H]
	\setlength{\abovecaptionskip}{0pt}
	\setlength{\belowcaptionskip}{0pt}
	\centering
		\includegraphics[height=8cm,width=12cm]{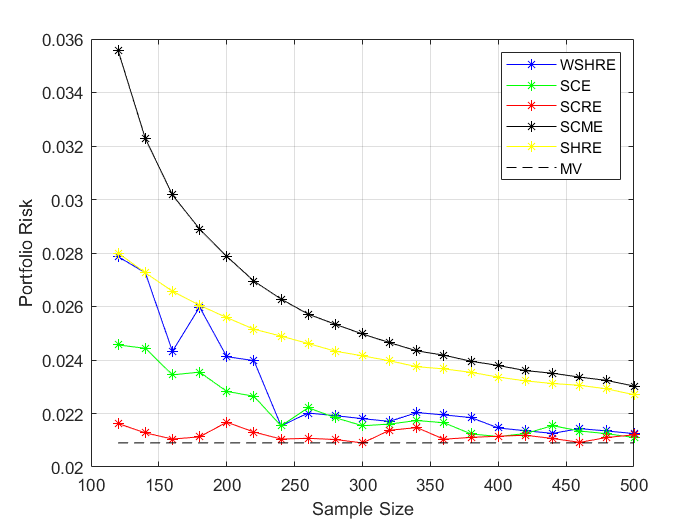}
	\caption{Portfolio variance vs. sample size $n$ for $M$. Comparison for SCME, SCRE, SCE, WSHRE, MV bound and SHRE with synthetic data.}
	\label{fig3}
\end{figure}

\subsection{Real Data}

For empirical analysis, we compare the performance of different optimal portfolio estimates by using real data from the S\&P 500 index. We download stock data for the S\&P 500 index from the website \href{a}{https://host.uniroma3.it/docenti/cesarone/DataSets.htm} from October 6, 2006 to December 31, 2020 and computer their weekly logarithmic returns. There are 336 stocks and 753 weekly observations for each stock. We randomly selected 100 stocks from 336 stocks in the S\&P 500 index. In the first comparison, select the top 200 as the sample set and 201 to 230 as the test set for prediction. For the second comparison, select 2 to 201 as the training set, 202 to 231 as the test set for prediction, and repeat 500 times in sequence.

To compare the performance of our proposed method SCRE with SCME, SHRE, WSHRE, and SCE. We use the following protocol for real datasets:

\begin{itemize}
\item Step 1: Select n weekly logarithmic returns as the training dataset and calculate the sample covariance. Spectral decomposition and correct the sample covariance under the spiked model, and estimate of spiked eigenvalues. Select the data from the next period of the training sample as the test dataset for estimating portfolio risk.
\item Step 2: Using the training dataset, derive the optimal parameters $\gamma^{*}_1$, $\gamma^{*}_2$ of the sample covariance of the SCRGMVP method using grid search over $\{\vp_1,\vp_2\}\in{\{[0,1)\times{[0,1)}\}}$. Design the SCRGMVP estimation and the other four estimators.
\item Step 3: Using the test dataset, estimate the portfolio risk for five methods.
\item Step 4: Repeat steps 1-3 500 times and calculate the average portfolio risk for all estimators.
\end{itemize}
The estimated portfolio risks using different methods are shown in the following figure:
\begin{figure}[H]
	\setlength{\abovecaptionskip}{0pt}
	\setlength{\belowcaptionskip}{0pt}
	\centering
		\includegraphics[height=8cm,width=12cm]{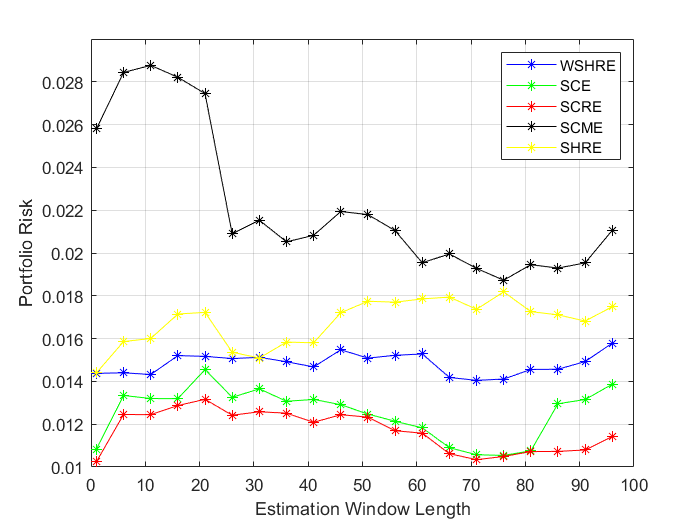}
	\caption{Portfolio variance vs. sample size $n$ for $M$. Comparison for SCME, SCRE, SCE, WSHRE, and SHRE with real data.}
	\label{fig1}
\end{figure}

\section{Conclusion}\label{conclu}
We propose a risk model that is superior to the classical model, namely the sample covariance matrix. Due to the shortcomings of the sample covariance matrix, we propose an enhanced alternative. Thus, the purpose of this paper is to solve the Markowitz optimization problem by developing a new covariance estimation to address the essence of portfolios. The key to this method is to assume that the population covariance matrix follows a spiked model. Using this special structure, the proposed estimator uses an estimation of the covariance matrix and follows the spiked model with regularized parameters. Using the results of the random theory, the asymptotic characterization of the proposed estimator is given, and the parameter that minimizes the consistent estimation of the estimator is selected. In this way, we can use the estimated eigenvalues and corresponding eigenvectors to preserve the form of the spiked covariance model as much as possible. Meanwhile, regularization methods can further reduce investment risks. Through the simulation of synthetic and real data, we found that our proposed method not only performs well compared to other methods but also greatly reduces the number of design parameters, which greatly reduces computational complexity. In further research, the same idea can also be applied to other backgrounds of Markowitz portfolio or other fields to improve the estimation of the population covariance matrix, such as linear discriminant analysis (LDA), Markowitz maximum return, etc.

\section*{Appendix A. Proof of Theorem \ref{thm:1}}\label{proof}
Firstly, according to (\ref{eq:angle}), we have
\begin{align}
	\fX\(\tfC\)&=\frac{\funit_M^T}{\sqrt{M}}\tfC\frac{\funit_M}{\sqrt{M}}\notag\\
	&=\frac{\funit_M^T}{\sqrt{M}}\left(\fI_M+\gamma_1\sum_{j\in\mathbb{I}_1}\lam_{j}\fu_j\fu_j^T+\gamma_2\sum_{j\in\mathbb{I}_2}\lam_{j}\fu_{j}\fu_{j}^T\right)^{-1}\frac{\funit_M}{\sqrt{M}}\notag\\
	&=\frac{\funit_M^T}{\sqrt{M}}\left(\fI_M-\sum_{j\in\mathbb{I}_1}\gamma_{1,j}\fu_j\fu_j^T-\sum_{j\in\mathbb{I}_2}\gamma_{2,j}\fu_{j}\fu_{j}^T\right)\frac{\funit_M}{\sqrt{M}}\notag\\
	&=1-\sum_{j\in\mathbb{I}_1}\gamma_{1,j}\frac{\funit_M^T}{\sqrt{M}}\fu_j\fu_j^T\frac{\funit_M}{\sqrt{M}}-\sum_{j\in\mathbb{I}_2}\gamma_{2,j}\frac{\funit_M^T}{\sqrt{M}}\fu_{j}\fu_{j}^T\frac{\funit_M}{\sqrt{M}}\notag\\
	&\xrightarrow{a.s.} 1-\sum_{j\in\mathbb{I}_1}\gamma_{1,j}a_jb_j-\sum_{j\in\mathbb{I}_2}\gamma_{2,j}a_{j}jb_{j}\notag\\
	&\triangleq \overline{X}\left(\gamma_1,\gamma_2\right),
\end{align}
in which
\beq
a_j = \frac{\lam_j^2-J}{\lam_j(\lam_j+J)},\;\;b_j = \(\frac{\funit_M^T}{\sqrt{M}}\fv_j\)^2,\notag
\eeq
\beq
\gamma_{1,j}&=&\frac{\vp_1\lam_j}{(1-\vp_1)\lam_1+\vp_1\lam_j},\;\;\;j \in \mathbb{I}_1,\\
\gamma_{2,j}&=&\frac{-\vp_2\lam_{j}}{(1-\vp_2)\lam_{-1}-\vp_2\lam_{j}},\;\;\;j \in \mathbb{I}_2.
\eeq
Secondly, rewrite \ref{eq:y}, we have 
\begin{align*}
	\fY(\tfC,\bSig)&=\frac{\funit_M^T}{\sqrt{M}}\tfC\bSig\tfC\frac{\funit_M}{\sqrt{M}}\notag\\
	&=\sigma_M^2\frac{\funit_M^T}{\sqrt{M}}\tfC\left(\fI_M+\sum_{j\in\mathbb{I}_1}\lam_{j}\fv_j\fv_j^T+\sum_{j\in\mathbb{I}_2}\lam_{j}\fv_{j}\fv_{j}^T\right)\tfC\frac{\funit_M}{\sqrt{M}}\notag\\
	&=\sigma_M^2\left(\frac{\funit_M^T}{\sqrt{M}}\tfC^2\frac{\funit_M}{\sqrt{M}}+\sum_{j\in\mathbb{I}_1}\lam_{j}\left(\frac{\funit_M^T}{\sqrt{M}}\tfC\fv_j)\right)^2+\sum_{j\in\mathbb{I}_2}\lam_{j}\left(\frac{\funit_M^T}{\sqrt{M}}\tfC\fv_{j}\right)^2\right).
\end{align*}\label{eq:y1}
For the three elements of \ref{eq:y1}, according to \ref{eq:angle}, we have
\begin{align*}
	\frac{\funit_M^T}{\sqrt{M}}\tfC^2\frac{\funit_M}{\sqrt{M}} \xrightarrow{a.s.} 	 1+\sum_{j\in\mathbb{I}_1}\left(\gamma_{1,j}^2-2\gamma_{1,j}\right)a_jb_j+\sum_{j\in\mathbb{I}_2}\left(\gamma_{2,j}^2-2\gamma_{2,j}\right)a_{j}b_{j},
\end{align*}
\begin{eqnarray*}
	&&\frac{\funit_M^T}{\sqrt{M}}\tfC\fv_j\fv_j^T\tfC\frac{\funit_M}{\sqrt{M}} \xrightarrow{a.s.} 	  b_j\(1-2\gamma_{i,j}a_j+\gamma_{i,j}^2a_j^2\).\\
\end{eqnarray*}

Similarly, for \ref{eq:y1}, we have
\begin{align*}
	\fY(\tfC,\bSig)&=\frac{\funit_M^T}{\sqrt{M}}\tfC\bSig\tfC\frac{\funit_M}{\sqrt{M}}\\
	&=\sigma_M^2\left(\frac{\funit_M^T}{\sqrt{M}}\tfC^2\frac{\funit_M}{\sqrt{M}}+\sum_{j\in\mathbb{I}_1}\lam_{j}\(\frac{\funit_M^T}{\sqrt{M}}\tfC\fv_j\)^2+\sum_{j\in\mathbb{I}_2}\lam_{j}\(\frac{\funit_M^T}{\sqrt{M}}\tfC\fv_{j}\)^2\right)\\
	&\xrightarrow{a.s.}\sigma_M^2\(\begin{array}{c}
		1+\sum_{j\in\mathbb{I}_1}\(\gamma_{1,j}^2-2\gamma_{1,j}\)a_jb_j\\
		+\sum_{j\in\mathbb{I}_2}\(\gamma_{2,j}^2-2\gamma_{2,j}\)a_{j}b_{j}\\
		+\sum_{j\in\mathbb{I}_1}\lam_{j}b_j\(1-2\gamma_{1,j}a_j+\gamma_{1,j}^2a_j^2\)\\
		+\sum_{j\in\mathbb{I}_2}\lam_{j}b_{j}\(1-2\gamma_{2,j}a_{j}+\gamma_{2,j}^2a_{j}^2\)
	\end{array}\)\\
	&=\sigma_M^2\(\begin{array}{c}
		1+\sum_{j\in\mathbb{I}_1}\lam_{j}b_j+\sum_{j\in\mathbb{I}_2}\lam_{j}b_{j}\\
		-2\sum_{j\in\mathbb{I}_1}\gamma_{1,j}(1+\lam_{j})a_jb_j\\
		-2\sum_{j\in\mathbb{I}_2}\gam _{2,j}(1+\lam_{j})a_{j}b_{j}\\
		+\sum_{j\in\mathbb{I}_1}\gamma_{1,j}^2(1+\lam_{j}a_j)a_jb_j\\
		+\sum_{j\in\mathbb{I}_2}\gamma_{2,j}^2(1+\lam_{j}a_{j})a_{j}b_{j}
	\end{array}\)\\
	&\triangleq\overline{Y}\left(\gamma_1,\gamma_2\right)
\end{align*}

\section*{Appendix B. Proof of Theorem \ref{thm:3}}

Let $\vp_i = \gamma_{i,1}\in(0,1)$ for $i=1,2$. Then, $\gamma_1$ and $\gamma_2$ can be rewritten as
\beqq
\gamma_1 = \frac{\vp_1}{(1-\vp_1)\lam_1}, \;\;\mbox{and}\;\;\gamma_2 = \frac{-\vp_2}{(1-\vp_2)\lam_{-1}}.
\eeqq
Further,
\beq
\gamma_{1,j}&=&\frac{\vp_1\lam_j}{(1-\vp_1)\lam_1+\vp_1\lam_j},\;\;\;j=1,2,...,r_1,\label{eq:ogamma1}\\
\gamma_{2,j}&=&\frac{-\vp_2\lam_{j}}{(1-\vp_2)\lam_{-1}-\vp_2\lam_{-j}},\;\;\;j=-1,-2,...,-r_2.\label{eq:ogamma2}
\eeq
Plugging (\ref{eq:ogamma1}) and (\ref{eq:ogamma2}) into (\ref{eq:oG}) and (\ref{eq:oD}), we have
\beqq
\widetilde{X}(\lam_1,\lam_2)=\overline{G}(\gamma_1,\gamma_2)\;\;\mbox{and}\;\;\widetilde{Y}(\lam_1,\lam_2)=\overline{D}(\gamma_1,\gamma_2).
\eeqq
Thus, the objective function can be written as
\beqq
	g(\vp_1,\vp_2)=\frac{\overline{Y}\left(\vp_1,\vp_2\right)}{\( \overline{X}\left(\vp_1,\vp_2\right)\)^2}
\eeqq
Thus, (\ref{eq:min0}) is equivalent to
\beqq
\vp_1^*,\vp_2^* = \mathop{\arg \min}\limits_{\vp_1, \vp_2\in(0,1)}g(\vp_1,\vp_2)
\eeqq
by which
\beqq
\gamma_1^* = \frac{\vp_1^*}{(1-\vp_1^*)\lam_1}, \;\;\mbox{and}\;\;\gamma_2^* = \frac{-\vp_2^*}{(1-\vp_2^*)\lam_{-1}}.
\eeqq

\bibliographystyle{unsrt}
\bibliography{referbib}

\end{document}